\documentclass[letterpaper,aps,10pt,superscriptaddress,twocolumn,floatfix,showpacs]{revtex4-1}
\usepackage{graphicx}
\usepackage{amsmath}
\usepackage{amsfonts}
\usepackage{float}
\usepackage{amssymb}
\usepackage{epsfig}
\usepackage{epstopdf}
\DeclareGraphicsExtensions{.pdf,.eps,.png,.jpg,.mps}
\usepackage[pdftex]{color}
\usepackage{amsmath,graphicx,amssymb,braket,xcolor,subfigure,upgreek}
\usepackage[colorlinks, linkcolor=blue, citecolor=blue, urlcolor=blue, breaklinks=true]{hyperref}
\usepackage{microtype}
\usepackage{bbm}
\usepackage{color}
\usepackage{dsfont}
\usepackage[english]{babel}
\usepackage{blindtext}
\usepackage[utf8]{inputenc}
\usepackage{textcomp}
\usepackage[title]{appendix}

\begin{document}

\title{Phase synchronization in dissipative non-Hermitian coupled quantum systems}
\author{Jonas Rohn}
\affiliation{Max Planck Institute for the Science of Light, Staudtstra{\ss}e 2,
D-91058 Erlangen, Germany}
\affiliation{Department of Physics, University of Erlangen-Nuremberg, Staudtstra{\ss}e 7,
D-91058 Erlangen, Germany}
\author{Kai Phillip Schmidt}
\affiliation{Department of Physics, University of Erlangen-Nuremberg, Staudtstra{\ss}e 7,
D-91058 Erlangen, Germany}
\author{Claudiu Genes}
\affiliation{Max Planck Institute for the Science of Light, Staudtstra{\ss}e 2,
D-91058 Erlangen, Germany}
\affiliation{Department of Physics, University of Erlangen-Nuremberg, Staudtstra{\ss}e 7,
D-91058 Erlangen, Germany}
\date{\today}

\begin{abstract}
We study the interplay between non-Hermitian dynamics and phase synchronization in a system of $\mathcal{N}$ bosonic modes coupled to an auxiliary mode.
The linearity of the evolution in such a system allows for the derivation of fully analytical results for synchronization conditions.
In contrast, analysis at the level of phase dynamics, followed by a transformation to a collective basis allows a complete reduction to an all-to-all coupled Kuramoto model with known analytical solutions.
We provide analytical and numerical solutions for systems ranging from a few modes to the macroscopic limit of large $\mathcal{N}$ in the presence of inhomogeneous frequency broadening and test the robustness of phase synchronization under the action of external noise.
\end{abstract}


\maketitle
Synchronization is a well-established field of research with wide relevance in a variety of disciplines ranging from biology to neuroscience, electrical engineering, mathematics and physics originating from Huygens' observation a few centuries back~\cite{huygens1893oeuvres, LIU1997855, schuster_biology, PhysRevLett.76.404, acebron2005thekuramoto, STROGATZ20001}.
In 1975 Kuramoto introduced the \textit{Kuramoto model}~\cite{kuramoto1984chemical,acebron2005thekuramoto, STROGATZ20001} to show the emergence of synchronization in multiple self-sustained oscillators with mutual, nonlinear couplings. Kuramoto discovered a clever analytical solution in the macroscopic limit of an infinite number of oscillators which provides an analytical estimate of the (identical) coupling strengths for the synchronization threshold.
The transition between the unsynchronized and the synchronized regime then is shown to resemble a second-order phase transition.
Nowadays, several variations of the original model are known, including the study of modified couplings as for example the Sakaguchi-Kuramoto model,  driving terms and other types of network topologies representing the mutual oscillator couplings \cite{10.1143/PTP.76.576, doi:10.1063/1.3049136, doi:10.1063/1.2930766}.

\begin{figure}[t]
	\centering
		\includegraphics[width=0.85\columnwidth]{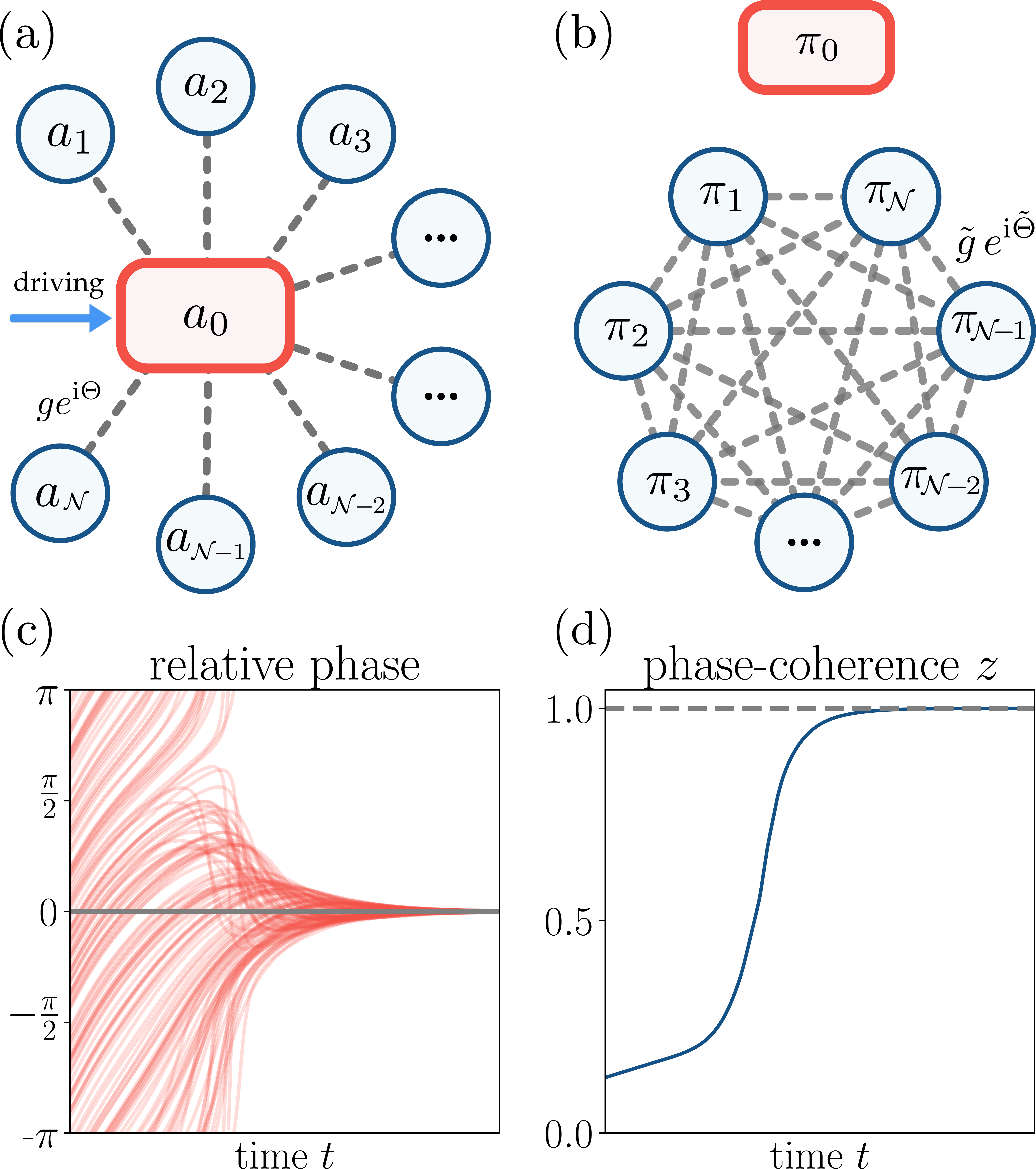}
		\caption{(a) Setup showing a (possibly) driven auxiliary mode $a_0$ identically coupled to $\mathcal{N}$ oscillators at rate $g e^{i\Theta}$. The tuning of $\theta$ allows for a smooth transition from Hermitian to fully anti-Hermitian dynamics. (b) A transformation into a collective \textit{Kuramoto basis} shows a resulting effective all-to-all coupling of $\mathcal{N}$ modes fully decoupled from an isolated, quickly decaying mode $\pi_0$. (c) and (d) Prototypical phase synchronization of all phases $\phi_j$ of \mbox{$\mathcal{N}=100$} oscillators coupled to the auxiliary mode in a fully anti-Hermitian fashion as a function of time $t$. The phase-coherence $z= \left| \sum_j \exp{(\mathrm{i}\phi_j)} \right|/\mathcal{N}$ is an alternative measure for synchronization where $z=1$ means full synchronization and $z=0$ indicates desynchronization.}
	\label{fig1}
\end{figure}

The emerging field of non-Hermitian physics recently aroused great interest as it offers an alternative formulation of quantum mechanics, but also is convenient  to describe open quantum systems with gain and loss ~\cite{bender1998nonHermitian, nonhermPTsymmetry2017, Rotter_2009, moiseyev_2011,RevModPhys.93.015005}. It is interesting to ask whether non-Hermitian interactions can be used as a resource for synchronization in the absence of an explicit non-linear ingredient. To this end, we consider a system of $\mathcal{N}$ oscillators \textit{linearly} coupled to a (possibly) driven auxiliary mode $a_0$ (see Fig.~\ref{fig1}a) with interactions smoothly tunable  from purely Hermitian to fully anti-Hermitian. We find that phase synchronization within such a system is reached both in the undriven (see Fig.~\ref{fig1}c, d) and driven case for identical oscillators. Most importantly, the non-Hermitian coupling is crucial, since optimal conditions for synchronization are obtained for fully anti-Hermitian systems while Hermitian systems only synchronize in the infinite-time limit. The linearity of the coupled system allows us to obtain fully analytical solutions.
%

For further insight into the nature of this synchronization, we provide an alternative analyis in the collective \textit{Kuramoto basis} obtained by a linear transformation of the original model. The \emph{Kuramoto modes} $\pi_j$ with $j=1,...\mathcal{N}$ exhibit an effective all-to-all coupling which can be exactly mapped onto the original Kuramoto model (see Fig.~\ref{fig1}b). This exact mapping allows to verify analytically that the synchronization conditions straightforwardly derived from the linear dynamics of the original non-Hermitian system coincide with the predictions of the Kuramoto model.\\
\indent Furthermore, we perform a numerical analysis of robustness against external amplitude noise captured within a stochastic Brownian noise model for a thermal bath. We find resilience to noise in driven systems and finite synchronization regimes for undriven models strongly dependent on the bath temperature. Additionally, we discuss the effect of frequency disorder stemming for example from inhomogeneous broadening; we find that low levels of disorder can be tolerated and that, past a certain threshold, the system synchronizes to two groups of opposing phases.\\

\noindent \textbf{Model} - We consider a subsystem of $\mathcal{N}$ identical bosonic \emph{main modes} $a_i$ at frequency $\omega$, identically coupled to the (possibly) driven auxiliary mode $a_0$ oscillating at frequency $\omega_0$. The effect of the environment is included via dissipation rates $\gamma$ and $\gamma_0$. Owing to the fluctuation-dissipation theorem, input noises stemming from the combined action of the bath modes will be included via operators $a_\mathrm{in}^{(i)}$, with $\braket{a_\mathrm{in}^{(i)}(t)}=0$ and $\braket{a_\mathrm{in}^{(i)}(t)a_\mathrm{in}^{(i)\dagger}(t')}=\delta(t-t')$ (all other correlations and cross-correlations vanish). The open system dynamics can then be described by the non-Hermitian Hamiltonian (see Appendix \ref{sec::APNDX1})
\begin{eqnarray}
	\mathcal{H}_\mathrm{NH} &=& (\omega_0 -\mathrm{i}\gamma_0) a_0^\dagger a_0^{\phantom{\dagger}} + \sum_{i=1}^\mathcal{N} (\omega -\mathrm{i}\gamma) a_i^\dagger a_i^{\phantom{\dagger}} \\
		&+& \frac{g}{\sqrt{\mathcal{N}}} e^{\mathrm{i}\Theta} \sum_{i=1}^\mathcal{N} (a_0^\dagger a_i^{\phantom{\dagger}} + a_0^{\phantom{\dagger}} a_i^\dagger) + \mathrm{i}\eta (a_0^\dagger e^{-\mathrm{i}\Omega t} - a_0 e^{\mathrm{i}\Omega t})\, . \nonumber
\end{eqnarray}
Here, $\eta \in \mathbb{R}$ is the driving strength of $a_0$, $\Omega$ the driving frequency and $g \in \mathbb{R}$ the coupling strength. Crucially, we allow the coupling between the main modes $i=1, \ldots, \mathcal{N}$
and the auxiliary mode $a_0$ to be non-Hermitian by introducing the
parameter $\Theta \in (-\pi, \pi]$. For $\Theta = 0$ or $\pi$ the usual case of a Hermitian, coherent interaction is obtained, characterized by the possible occurrence of strong coupling physics whereas $\Theta =\pm \pi/2$ yields a fully anti-Hermitian coupling term leading to level attraction dynamics.

\indent The evolution of the system is given by the quantum Langevin equations of motion (QLEs) \mbox{$\dot{a}_i = \mathrm{i} [\mathcal{H}_\mathrm{NH}, a_i] - \sqrt{2\gamma_i} a_\mathrm{in}^{(i)}$} written as generalizations of the Heisenberg equations of motion to include dissipation and fluctuations~(see Appendix \ref{sec::APNDX1}).
Next we focus on the predicted evolution of the expectation values $\alpha_i := \langle a_i \rangle \in \mathbb{C}$
while later we will discuss the effects of the noise terms. Explicitly writing out the equations of motion for all $\alpha_i$ yields (in a frame rotating at the laser frequency $\Omega$) a set of
$\mathcal{N}+1$ linear, coupled, complex differential equations
\begin{equation}
\label{eq::compl_ampl_eom}
  \dot{\mathbf{A}} = -\mathrm{i} \underbrace{\left(
  \begin{array}{cccc}
  \delta_0 - \mathrm{i} \gamma_0 & \frac{g}{\sqrt{\mathcal{N}}} e^{\mathrm{i}\Theta} & \ldots & \frac{g}{\sqrt{\mathcal{N}}}e^{\mathrm{i}\Theta} \\
  \frac{g}{\sqrt{\mathcal{N}}} e^{\mathrm{i}\Theta} & \delta - \mathrm{i} \gamma &  &    \\
  \vdots &  & \ddots &  \\
  \frac{g}{\sqrt{\mathcal{N}}} e^{\mathrm{i}\Theta} &  &  & \delta - \mathrm{i} \gamma
  \end{array}\right)}_\mathcal{H} \mathbf{A} + \mathbf{\eta}\, \mathbf{u}
\end{equation}
with the evolution matrix $\mathcal{H}$, \mbox{$\mathbf{A} = (\alpha_0, \alpha_1, \ldots, \alpha_\mathcal{N})^T$}, the driving vector \mbox{$\mathbf{u} := (1, 0, \ldots, 0)^T$}, and the frequency detunings $\delta_{0} = \omega_{0} - \Omega$, $\delta = \omega - \Omega$. \\

\noindent \textbf{Analytical results} -  Let us now provide analytical results for synchronization conditions based solely on the linear evolution from Eq.~\eqref{eq::compl_ampl_eom}. In order to analytically and numerically insure that the phase difference between two oscillators is $0$ or $\pi$ in the long-time limit, we require $\left.\Im{(\alpha_i/\alpha_j)}\right|_{t \rightarrow \infty} = 0$. To further distinguish between synchronization (zero phase difference) and anti-synchronization (phase difference of $\pi$) we further require $\left.\Re{(\alpha_i/\alpha_j)}\right|_{t \rightarrow \infty} > 0$. The task at hand is to find values of the non-Hermitian angle parameter $\Theta$ and driving frequency $\Omega$ that lead to phase synchronization of all oscillators, assuming that $\omega_{0}, \gamma_{0}, \omega, \gamma, g$ and $\eta$ are given.
In the case where external driving is absent \mbox{($\eta = 0$)}, as amplitudes eventually decay to zero in the steady state long-time limit, we will consider synchronization to be reached only when the synchronization time $\tau_\mathrm{sync}$ is smaller than the decay time $\tau_\mathrm{dec}$.
To calculate the ratios $\alpha_i/\alpha_j$, Eq.~\eqref{eq::compl_ampl_eom} is solved analytically by diagonalizing $\mathcal{H}$ (see Appendix \ref{sec::APNDX2}).
Note that this approach is not applicable for an arbitrary choice of parameters $\omega_0, \gamma_0, \omega, \gamma, g, \Theta$ since $\mathcal{H}$ is non-Hermitian and therefore not diagonalizable in general. In practice however, this special case is still covered by the outcomes obtained assuming $\mathcal{H}$ to be diagonalizable.

More importantly, we have to distinguish the driven case $\eta \ne 0$ and the undriven case $\eta = 0$ as the resulting synchronization conditions are different.\\

\begin{figure}
 \centering
  \includegraphics[width=0.95\columnwidth]{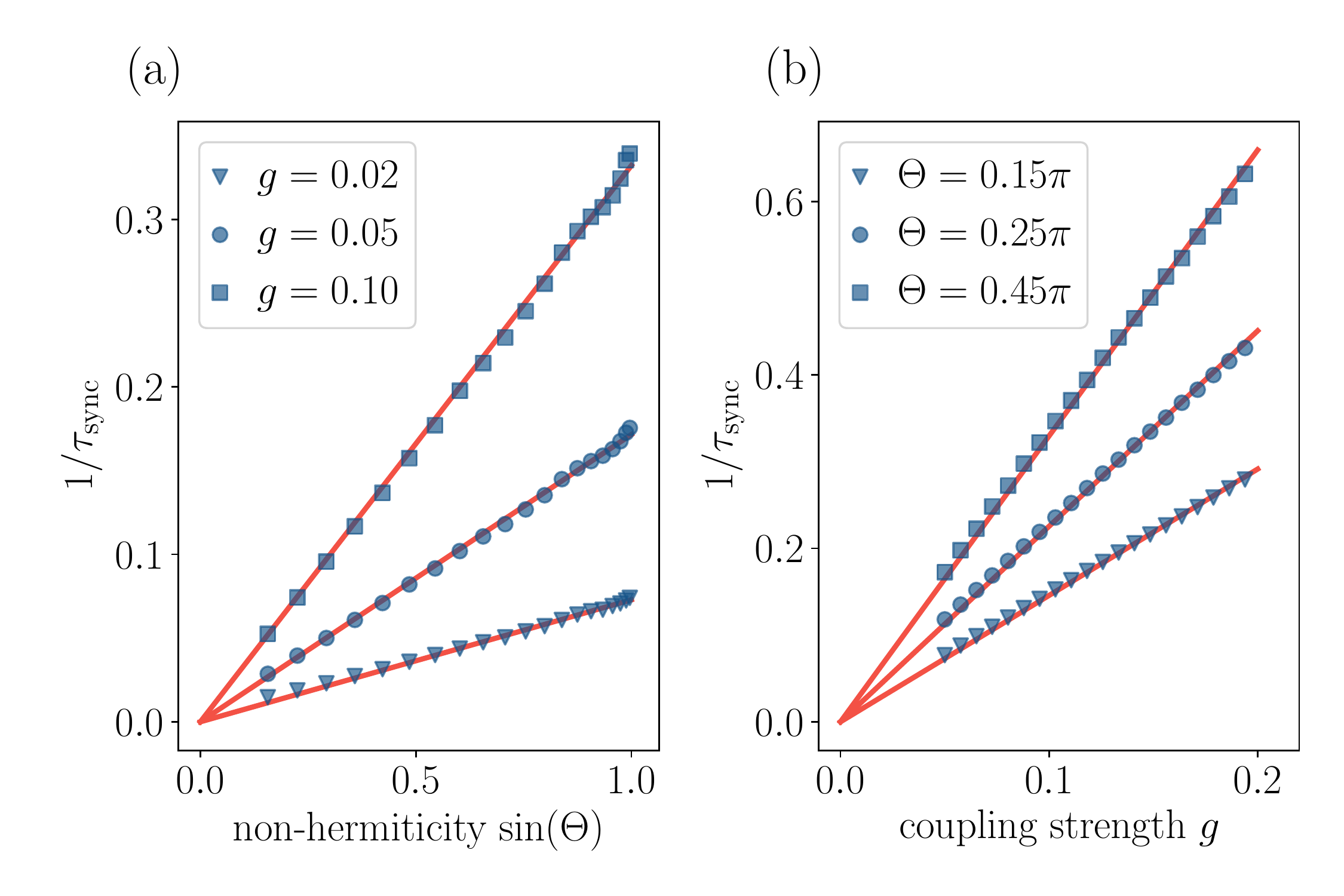}
  \caption{Inverse mean synchronization time $\tau_\mathrm{sync}^{-1}$ as a function of (a) the non-Hermitian parameter $\sin(\Theta)$ and (b) the coupling strength $g$. Here, $\tau_\mathrm{sync}$ is numerically estimated by averaging the time by which phases deviate from the auxiliary mode phase $\phi_0$ not more than $0.05\pi$. Linear lines are fits through the data points shown as symbols.}
  \label{fig::sync_time}
\end{figure}

\noindent \emph{Undriven case} - For $\eta = 0$, \mbox{$g>0$}, all amplitudes $\alpha_i(t)$ are superpositions of eigenmodes which will eventually decay.
However, since one single eigenmode survives the other ones, for $t \rightarrow \infty$ the ratios $\alpha_i / \alpha_j$ become independent of the initial amplitudes and approach a constant value which is closely related to entries of the corresponding eigenvector.
One can then show (see Appendix \ref{sec::APNDX2}) that synchronization is achieved if
\begin{equation}
\label{eq::sync_cond_undriven}
  \begin{aligned}
  	\tan{\Theta} = -\frac{\gamma - \gamma_0}{\omega - \omega_0} = -\frac{\Delta \gamma}{\Delta \omega}
  \end{aligned}
\end{equation}
(stemming from the condition $\left.\Im{(\alpha_i/\alpha_j)}\right|_{t \rightarrow \infty} = 0$) combined with the condition that $\Delta \gamma \sin{\Theta} < \Delta \omega \cos{\Theta}$ (stemming from the requirement $\left.\Re{(\alpha_i/\alpha_j)}\right|_{t \rightarrow \infty} > 0$).
Interestingly, the strength of the interaction $g$ does not play a role whereas it is crucial that the interaction is non-Hermitian.
 In particular, for \mbox{$\Delta \omega = 0$}, \mbox{$\Delta\gamma < 0$}, one finds $\Theta = \pi / 2$, demonstrating the necessity of non-Hermitian interactions.
 Surprisingly, the Hermitian case $\Theta = 0$ which could be obtained by choosing $\Delta \gamma = 0$, $\Delta \omega > 0$ practically does not lead to synchronization since the synchronization time $\tau_\mathrm{sync}$ diverges cf. Fig. \ref{fig::sync_time}.
 Calling the system synchronized is only sensible unless  $\tau_\mathrm{sync}$ does not exceed the decay time $\tau_\mathrm{dec}$ of the system.
 We find numerically that $\tau_\mathrm{sync}^{-1}$ scales as $g\sin{\Theta}$ whereas $\tau_\mathrm{dec}=1/\gamma$ (see Fig. \ref{fig::sync_time}).\\

\noindent \emph{Driven case} - The situation is different for a finite driving $\eta > 0$, since the system will eventually reach the steady state where all power from the driving is dissipated at equal rate.
The oscillators thereby assume the driving frequency $\Omega$, but their mutual phase differences do not vanish in general.
However, by choosing proper $\Theta, \Omega$ so that the synchronization condition
\begin{equation}
\label{eq::sync_cond_driven}
	\tan{\Theta} = -\frac{\gamma}{\omega - \Omega}\quad ,
\end{equation}
combined with $(\omega - \Omega) \, \cos{\Theta} < \gamma\, \sin{\Theta}$, are fulfilled, all phases coincide as soon as the oscillators enter the steady state.
The driving strength $\eta$ determines the steady state amplitudes and the synchronization time $\tau_\mathrm{sync}$, but similar to $g$, it does not appear in the synchronization condition.
Like in the undriven case, a non-Hermitian coupling is required in general, but $\tau_\mathrm{sync}$ may not be finite.
One finds numerically, that the synchronization time diverges in the Hermitian case $\Theta = 0$ as for the undriven case.\\

\noindent \textbf{Phase evolution} -  While Eqs.~(\ref{eq::compl_ampl_eom}) show linear evolution at the level of complex-valued amplitudes, the dynamics of phases \mbox{$\phi_i = \arg{\alpha_i}$} and real-valued amplitudes \mbox{$r_i = |\alpha_i|$} is described by a coupled system of $2(\mathcal{N}+1)$ nonlinear differential equations. With concise notations for coupling rates $\zeta_{0, j}=1-\delta_{0,j}$, $\zeta_{i>0, j}=\delta_{i, j}$, frequency detunings $\delta_i=\delta_{i, 0}\delta_0 + (1-\delta_{i,0})\delta$ and decay rates are defined in simplified form as $\gamma_i=\delta_{i, 0}\gamma_0 + (1-\delta_{i,0})\gamma$, the evolution is described by
\begin{equation}
\label{eq::ampl_phase eom}
  \begin{aligned}
  \dot{r}_i &= -\gamma_i r_i + \frac{g}{\sqrt{\mathcal{N}}} \sum_{j=1}^\mathcal{N} \zeta_{i, j} \sin{(\Theta + \phi_j - \phi_i)} r_j + \eta \cos{\phi_i} \\
  	\dot{\phi}_i &= - \delta_i - \frac{g}{\sqrt{\mathcal{N}}} \sum_{j=1}^\mathcal{N} \zeta_{i, j} \frac{r_j}{r_i} \, \cos{(\Theta + \phi_j - \phi_i)} - \eta \sin{\phi_i}.
  \end{aligned}
\end{equation}
Assuming that the initial phases $\phi_j$ at $t=0$ are randomly distributed in the interval $0$ and $2\pi$, synchronization behaviour is found if all phase differences approach zero, i.~e., if \mbox{$\phi_i(t) \approx \phi_j(t), \forall i, j \in \{0, \ldots, \mathcal{N}\}$ for times $t>\tau_\mathrm{sync}$}, where $\tau_\mathrm{sync}$ denotes an estimate of the synchronization time.\\
\indent Alternatively, we quantify synchronization via the phase-coherence $z \equiv \left| \sum_i e^{\mathrm{i}\phi_i} /\mathcal{N}\right|$ which reaches unity for full synchronization and is bounded $0 \le z <1$ when some of the oscillators are not fully synchronized.\\
\indent The form of Eqs.~\eqref{eq::ampl_phase eom} already hints at synchronization as the phase equations are remarkably similar to variations of the well-known Kuramoto model \cite{kuramoto1984chemical}.
Indeed, if $r_j/r_i \approx \mathrm{const.}$ (as numerical simulations show in the long-time limit) the phase equations completely decouple from the amplitude equations. The obtained model features the same interaction term as the Sakaguchi-Kuramoto model \cite{10.1143/PTP.76.576}, thus being a driven variation for a star-like topology as shown in Fig.~\ref{fig1}a. However, we will see that an exact mapping onto the original, analytically solvable Kuramoto model with all-to-all coupling can only be obtained in a collective basis where only $\mathcal{N}$ modes are synchronized.\\

\noindent \textbf{Mapping onto the Kuramoto model}\label{sec::kuramoto_mapping} -- The main ingredient of the Kuramoto model is the all-to-all interaction among all participants in the synchronization process. This is at a difference with the structure of the matrix $\mathcal{H}$ in Eq.~\eqref{eq::compl_ampl_eom} where the $\mathcal{N}$ modes are coupled to the single auxiliary mode but not to each other. The all-to-all coupling limit can however be reached within the reduced subspace of dimension $\mathcal{N}$ comprised of a set of collective modes obtained by the action of linear mapping $\mathcal{U}$. This \textit{collective basis} consists of $\mathcal{N}$ Kuramoto modes, equally coupled to each other and an isolated eigenmode (see illustration in Fig.~\ref{fig1}b).\\
\indent In this basis, the equation of motion for the transformed complex amplitudes $\mathbf{P} = (\pi_0, \pi_1, \ldots, \pi_\mathcal{N})^T = \mathcal{U} \mathbf{A}$ reads $\dot{\mathbf{P}} = -\mathrm{i} \mathcal{M} \mathbf{P} + \eta \tilde{\mathbf{u}}$ with an effective driving vector $\tilde{\mathbf{u}} = \mathcal{U}\mathbf{u}$ and the  evolution matrix $\mathcal{M} = \mathcal{U H U}^{-1}$ is explicitly given by
\begin{equation}
   \mathcal{M} = \left(
  	\begin{array}{cccc}
  	\tilde{\delta}_0 - \mathrm{i} \tilde{\gamma}_0 & 0 & 0 & \ldots \\
  	0 & \tilde{\delta} - \mathrm{i} \tilde{\gamma} & \frac{\tilde{g}}{\mathcal{N}} e^{\mathrm{i}\tilde{\Theta}} &  \ldots   \\
  	0 & \frac{\tilde{g}}{\mathcal{N}} e^{\mathrm{i}\tilde{\Theta}} &  \tilde{\delta} - \mathrm{i} \tilde{\gamma} & \\
  \vdots & \vdots &  & \ddots
 \end{array}\right) \quad .\end{equation}
The renormalized couplings are obtained as
\begin{equation}
\tilde{g} e^{\mathrm{i}\tilde{\Theta}} = \frac{1}{2}(\Delta \omega -\mathrm{i} \Delta \gamma \mp 2\mu)
\end{equation}
with $\mu = \sqrt{\left(\Delta \omega - \mathrm{i}\Delta \gamma\right)^2/4 + g^2 e^{2\mathrm{i}\Theta}}$.
The detunings and decay rates are also redefined as
\begin{subequations}
\begin{align}
   \tilde{\delta}_0 - \mathrm{i} \tilde{\gamma}_0 &= ((\delta + \delta_0) -\mathrm{i} (\gamma + \gamma_0))/2 \pm \mu, \\
    \tilde{\delta} - \mathrm{i} \tilde{\gamma} &= \delta - \mathrm{i} \gamma + (\Delta \omega - \mathrm{i} \Delta \gamma  \mp 2\mu)/({2\mathcal{N}})\, .
\end{align}
\end{subequations}
The $\pm$ signs appearing in the definition of the effective parameters originate from the non-uniqueness of the transformation $\mathcal{U}$.
In the following we will assume that $\mathcal{U}$ always isolates the eigenmode with the fastest decay rate, which is always possible.
\begin{figure}
 \centering
  \includegraphics[width=0.68\columnwidth]{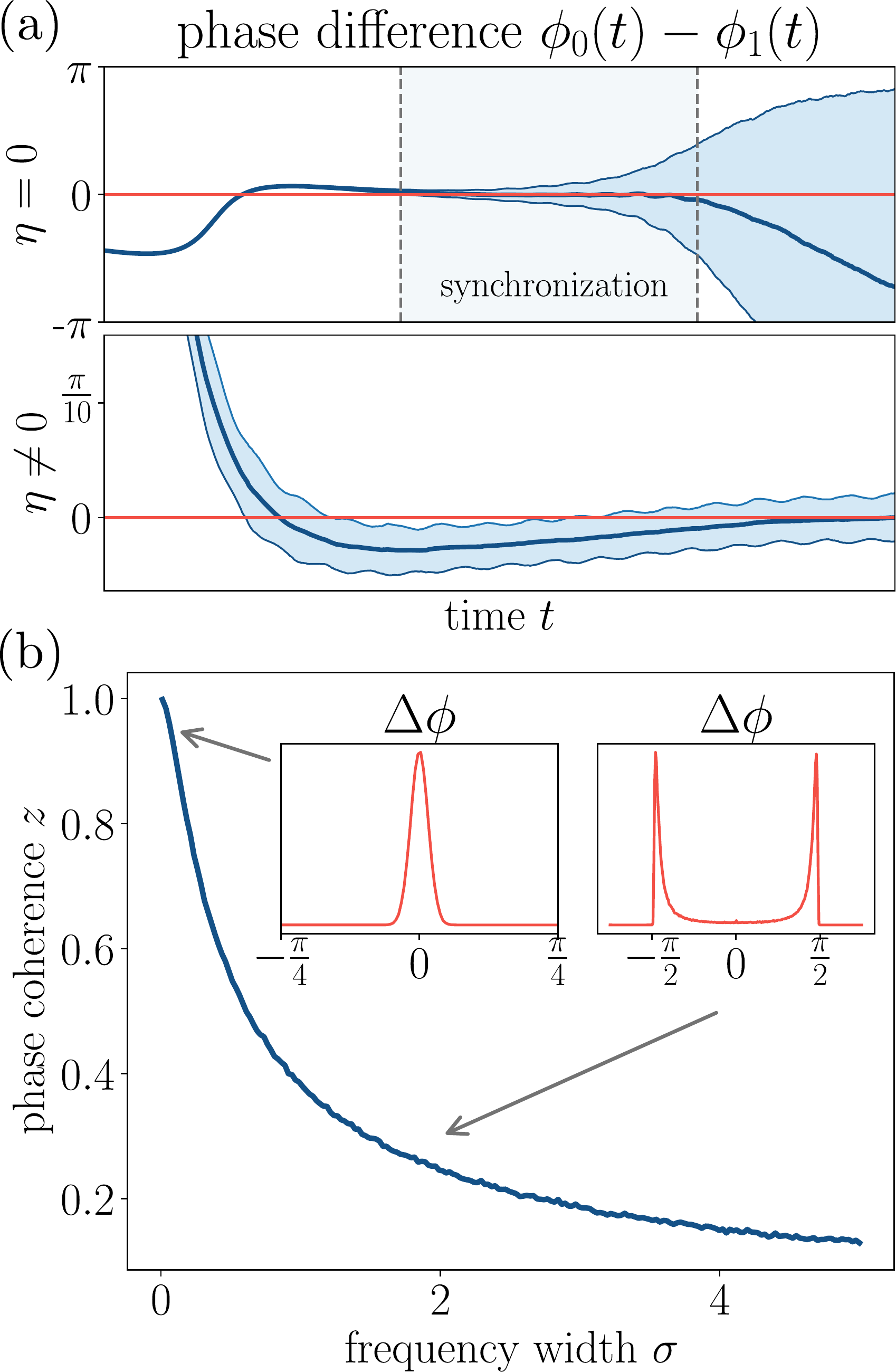}
  \caption{(a) Phase difference between two coupled oscillators with additional noise. In the undriven case $\eta = 0$, the oscillators synchronize, but the variance increases linearly at some time and the mean value deviates from zero. For the driven system $\eta \ne 0$, the variance and mean value remain finite and zero, respectively.
  		(b) Phase coherence as a function of the frequency disorder $\sigma$. For small $\sigma$ the distribution of phase differences is approximately Gaussian, and thus $z \approx 1$. In contrast, for $\Theta = \frac{\pi}{2}$ the phase distribution approaches a balanced bimodal distribution where the peaks are located at $\pm \pi/2$. Therefore the phase coherence $z \rightarrow 0$ for $\sigma \rightarrow \infty$ since the width of both peaks decreases with increasing disorder.}
  \label{fig::freq_disorder}
\end{figure}
With the phases $\psi_i := \arg{\pi_i}$ and real-valued amplitudes $\rho_i = |\pi_i|$ in the collective basis, one finds that the phase of the uncoupled mode $\psi_0$ increases linearly with time.
The phase evolution of the Kuramoto modes instead is
\begin{equation}
\label{eq::kuramoto_all_to_all}
  \dot{\psi}_i = -\tilde{\delta} - \frac{\tilde{g}}{\mathcal{N}} \sum_{\substack{j=1 \\ j\neq i}}^\mathcal{N} \cos{(\tilde{\Theta} + \psi_j - \psi_i)} \frac{\rho_j}{\rho_i} - \frac{\tilde{\eta}}{\rho_i} \sin{(\Theta_D - \psi_i)},
\end{equation}
with the renormalized driving strength $\tilde{\eta}$ and  phase delay $\Theta_D$ given by $\tilde{\eta} e^{\mathrm{i} \Theta_D} = \eta \frac{g e^{\mathrm{i} \Theta}}{\frac{\Delta \omega - \mathrm{i}\Delta \gamma}{2} \pm \mu}$.
If at least in the long-time limit  $\lim_{t \rightarrow \infty}\rho_j/\rho_i \rightarrow 1$, Eq.~ (\ref{eq::kuramoto_all_to_all}) resembles a class of all-to-all coupled Kuramoto model \cite{acebron2005thekuramoto}.

\indent For an undriven system $\eta = 0$, one can show that the complex amplitudes approach the same value if the isolated eigenmode is decaying faster than the other eigenmodes since then the transformation $\mathcal{U}$ mixes the long-living eigenmode at equal weight into all $\pi_i$ with $i\in\{1,\ldots,\mathcal{N}\}$.
For $\eta > 0$, the amplitudes $\rho_i$ reach the same steady state since the effective driving as well as the decay rates are identical in the collective basis (see Appendix \ref{sec::APNDX2}).
Hence, one indeed finds that $\lim_{t \rightarrow \infty}\rho_j/\rho_i\rightarrow 1$ and the Sakaguchi-Kuramoto model with driving and all-to-all coupling is obtained.
Due to the frequency distribution of the form $\delta(\omega - \omega_\mathrm{eff})$, the system can synchronize for any coupling strength $\tilde{g} > 0$ independent of $\tilde{\Theta}$ and $\eta$ \cite{10.1143/PTP.76.576, doi:10.1063/1.2930766}.
Therefore, all Kuramoto-modes $i=1, \ldots, \mathcal{N}$ are phase synchronized, in contrast to $\pi_0$.
In the undriven case, the phase of the Kuramoto modes and the eigenmode are unrelated and synchronization of the modes $\alpha_1, \ldots, \alpha_\mathcal{N}$ in the bare basis is achieved since the back-transformation $\mathcal{U}^{-1}$ treats these modes identically (see Appendix \ref{sec::APNDX2}).
Full synchronization, meaning that also the auxiliary mode $\alpha_0$ attains the same phase as the other modes, is obtained if additionally the condition stated in Eq.~\eqref{eq::sync_cond_undriven} is satisfied.
For $\eta > 0$, the phases of all oscillators in the Kuramoto-basis are equal for $t > \tau_\mathrm{sync}$ and synchronization of all modes in the bare bases leads to the synchronization condition \eqref{eq::sync_cond_driven}.\\
\indent For both, the driven and the undriven cases, the mechanism of synchronization of the main modes $i=1, \ldots, \mathcal{N}$ can therefore be led back to derivations of Kuramoto models which are exactly solvable.
If additionally the conditions \eqref{eq::sync_cond_undriven} (for undriven case) or \eqref{eq::sync_cond_driven} (for the driven case) are fulfilled,
all phases including that of the auxiliary mode approach the same steady-state phase consistent with results previously highlighted.\\

\noindent \textbf{Discussions} -- Two fundamental aspects can strongly perturb synchronization: the effect of external noise and the inherent frequency disorder in the system. In order to tackle the first aspect, we consider the stochastic equations of motion for the complex amplitudes $\alpha_i$ in the presence of thermal noise
\begin{equation}
\label{eq::complex_eom_sde}
  d\mathbf{A} = \left(-\mathrm{i} \mathcal{H} \mathbf{A} + \eta \mathbf{u} \right) dt + \mathrm{i} d\mathbf{W},
\end{equation}
where $d\mathbf{W} = (dW_0, dW_1, \ldots, dW_\mathcal{N})^T$ denotes a vector of $\mathcal{N}+1$ zero-averaged independent Wiener increments with \mbox{$\langle dW_i \rangle = 0$} and correlations $\langle dW_i dW_j \rangle = \delta_{i, j} \, \xi_i \xi_j dt$~ (see Appendix \ref{sec::APNDX3}). The noise entering the system comes from a bath at temperature $T$ which enters in the weights \mbox{$\xi_i = \sqrt{2 \gamma_i n_i(T)}$} depending on the loss rates \mbox{$\gamma_{i=0} = \gamma_0, \gamma_{i>0}=\gamma$} and the average occupation number $n_i = k_B T/\hbar \omega_i$, with the frequencies $\omega_{i=0}=0, \omega_{i>0}=\omega$.
We solve Eq.~\eqref{eq::complex_eom_sde} numerically by drawing random sample paths.
To test the resilience of the synchronized system against thermal noise we numerically follow the time evolution of the variance of the phase.\\
\indent In the undriven case $\eta = 0$, synchronization is preserved as long as the amplitude of each oscillator is larger than the fluctuations imposed by the random force stemming from the thermal bath.
The system is then only synchronized for $\tau_\mathrm{sync} < t < \tau_\mathrm{noise}$, where $\tau_\mathrm{noise}$ denotes the time by which the system is dominated by thermal noise.
Numerical results (see Fig.~\ref{fig::freq_disorder}a) show that the variance of the mean phase diverges for $t \rightarrow \infty$. For finite driving $\eta > 0$, the system always remains synchronized as the variance is bounded.
Therefore the phase difference with respect to the auxiliary mode cannot grow arbitrarily large and stays close to the average value (see Fig.~\ref{fig::freq_disorder}a).

\indent In order to tackle the question of frequency disorder, we allow the frequencies of the modes $i = 1, \ldots, \mathcal{N}$ to be disordered according to a  symmetric frequency distribution $g(\omega)$ peaked around the mean frequency $\bar{\omega}$.
Numerical results for a Gaussian distribution of the form $\exp{(-(\omega - \bar{\omega})^2/2\sigma^2)}$ suggest that the synchronization conditions hold for small $\sigma \ll \bar{\omega}$, but the steady-state phase difference $\Delta\phi = \phi_0 - \phi_i$ is  distributed normally with a modified variance $\sigma_\phi$ depending on the interaction strength, the width of the frequency disorder distribution $\sigma$ and the non-Hermitian parameter $\Theta$ (see Fig.~\ref{fig::freq_disorder}b).
Interestingly, $\sigma_\phi$ is proportional to $\sigma$, i.~e., $\sigma_\phi = c(g, \Theta) \sigma$ and $c(g, \Theta)$ is largest at $\Theta = \pi/2$.
For arbitrarily large $\sigma$ however, the distribution becomes more complex, and in the limit $\sigma \gg \bar{\omega}$ a Chimera-like distribution is approached, where the population of oscillators is split into two groups.
In the case $\Theta = \pi/2$, the distribution is symmetric around $\Delta \pi = 0$ with the two peaks at $\pm \pi/2$.
Since the peaks become the narrower the larger $\sigma$ is, the phase coherence $z$ can be used to distinguish different regimes.
 Weak disorder leads to a nearly synchronized population, and thus $z \approx 1$.
 The more the distribution separates into two sub-groups, the smaller is $z$ which converges to $0$ algebraically.
 In principle, one could again apply the formalism presented in the last section using a transformation into the Kuramoto basis.
 However, due to non-uniform frequencies, not only the eigenmode and the Kuramoto mode are coupled, also the all-to-all coupling now has a random strength which complicates analytical considerations drastically.
 Therefore, the complete investigation of this question needs further research.\\
%

\textbf{Acknowledgments} -- We acknowledge useful discussions with Muhammad Asjad and Christian Sommer. We acknowledge financial support from the Max Planck Society and from the German Federal Ministry of Education and Research, co-funded by the European Commission (project RouTe), project number 13N14839 within the research program "Photonik Forschung Deutschland" (C.~G.). We further acknowledge support by the Deutsche Forschungsgemeinschaft (DFG, German Research Foundation) -- Project-ID 429529648 -- TRR 306 QuCoLiMa ("Quantum Cooperativity of Light and Matter’’).

\bibliography{Refs}
\onecolumngrid
\appendix

\section{Non-Hermitian dynamics}
\label{sec::APNDX1}

\subsection{From Hermitian open system to closed system non-Hermitian dynamics}
In order to justify the non-Hermitian equations of motion for the expectation values of $\mathcal{N}+1$ coupled modes we first derive so-called Langevin equations for a Hamiltonian consisting of three parts $H = H_\mathrm{sys} + H_\mathrm{bath} + H_\mathrm{int}$
\begin{equation}
  \begin{aligned}
  		H_\mathrm{sys} &= \sum_{i=0}^\mathcal{N} \omega_i a_i^\dagger a_i + \sum_{i \neq j} g_{i, j} (a_i^\dagger a_j + a_i a_j^\dagger) + \underbrace{\sum_{i=0}^\mathcal{N} \mathrm{i}\eta_i (a_i^\dagger e^{-\mathrm{i}\Omega_i t} - a_i e^{+\mathrm{i}\Omega_i t})}_{H_\mathrm{drive}}\\
  		H_\mathrm{bath} &= \sum_{i=0}^\mathcal{N} \int d\omega \omega b_i^\dagger(\omega)b_i(\omega) \\
  		H_\mathrm{int} &= \sum_{i=0}^\mathcal{N} \int d\omega k_i(\omega) (b_i^\dagger(\omega) a_i + b_i(\omega) a_i^\dagger)
  \end{aligned}
\end{equation}
Here, $H_\mathrm{sys}$ denotes the Hamiltonian of the system, which includes $\mathcal{N}+1$ modes which interact mutually and are coherently driven where $\Omega_i$ are the driving frequencies.
Each mode additionally is coupled to an external bath of modes (Hamiltonian $H_\mathrm{bath}$ and $H_\mathrm{int}$ for the interaction) representing the environment into which the system modes can dissipate energy.
A common procedure to eliminate the bath modes is to change to the Heisenberg picture and insert the formally integrated equation of motion for the bath operators into the equation of motion for the operators $a_i$.
First, the Heisenberg equations of motion are
\begin{equation}
  \begin{aligned}
  	\dot{a}_i &= \mathrm{i} [H_\mathrm{sys}, a_i] - \mathrm{i} \int d\omega k_i(\omega) b_i(\omega) \\
  	\dot{b}_i(\omega) &= -\mathrm{i} \omega b_i(\omega) - \mathrm{i} k_i(\omega) a_i,
  \end{aligned}
\end{equation}
and hence
\begin{equation}
  b_i(\omega) = e^{-\mathrm{i} \omega (t - t_0)} b_i(\omega) - \mathrm{i} k_i(\omega) \int_{t_0}^t \mathrm{d}t^\prime e^{-\mathrm{i}\omega (t - t^\prime)} a_i,
\end{equation}
which is used to eliminate $b_i$ in the equations for $a_i$.
Assuming $k_i^2(\omega) = \gamma_i / \pi = \mathrm{const.}$, i. e. making the Markov approximation, we obtain the Langevin equations
\begin{equation}
  \dot{a}_i = \mathrm{i} [H_\mathrm{sys}, a_i]  - \gamma_i a_i(t) - \sqrt{2\gamma_i} a_\mathrm{in}^{(i)}.
 \end{equation}
$a_\mathrm{in}^{(i)} = \frac{\mathrm{i}}{\sqrt{2\pi}} \int d\omega e^{-\mathrm{i}\omega (t-t^\prime)} b_i(\omega)$ is an input operator hiding the back action of the bath onto the system.
Including the term $\gamma_i a_i(t)$ in the system Hamiltonian in a way, such that it is produced by the commutator in the Langevin equation allows us to define a non-Hermitian Hamiltonian $H_\mathrm{NH} = H_\mathrm{sys} - \sum_{i}\mathrm{i}\gamma_i a_i^\dagger a_i$.
One can now imagine that for example by choosing another basis, the non-Hermitian Hamiltonian can be transformed to another Hamiltonian which has a non-Hermitian interaction term.
Alternatively, as shown in the next section, non-Hermitian interactions can be also seen as simplified, effective descriptions of coupled dissipative modes.
In any case, we can generalize the formalism of Langevin equations by permitting any non-Hermitian Hamiltonian $H_\mathrm{HN}$ of the form $H_\mathrm{NH} = \sum_i \tilde{\omega}_i a_i^\dagger a_i + \sum_{i \neq j} g_{i, j} (a_i^\dagger a_j + a_i a_j^\dagger) + H_\mathrm{drive}$ with $\tilde{\omega}_i, g_{i, j} \in \mathbb{C}$.
The Langevin equations then read
\begin{equation}
\label{eq:nh_langevin}
  \dot{a}_i = \mathrm{i} [H_\mathrm{NH}, a_i] - \sqrt{2\gamma_i} a_\mathrm{in}^{(i)}.
\end{equation}

\subsection{Non-Hermitian interactions}

Here, we present an approach to implement effectively non-Hermitian interactions, although there are several other possibilities.
Let us first consider two modes which are indirectly coupled by interacting with a common auxiliary mode.
Then, the corresponding Langevin equations are given by Eq.~(\ref{eq:nh_langevin}) with the non-Hermitian Hamiltonian
\begin{equation}
  H_\mathrm{NH}^0 = (\omega_1 - \mathrm{i}\gamma_1) a_1^\dagger a_1 + (\omega_2 - \mathrm{i}\gamma_2) a_2^\dagger a_2 + (\Omega -\mathrm{i} \Gamma) \Lambda^\dagger \Lambda + ( g_1 a_1^\dagger + g_2 a_2^\dagger ) \Lambda + \mathrm{h. c.}\quad ,
\end{equation}
where $\omega_1, \omega_2, \Omega$ and $\gamma_1, \gamma_2, \Gamma$ denote frequencies and decay rates of mode $a_1, a_2$ and the auxiliary mode $\Lambda$.
The coupling strengths $g_1, g_2$ are allowed to be complex, but the total interaction term is still defined to be Hermitian.
Explicitly writing out the Langevin equations for the expectation values $\alpha_i := \langle a_i \rangle$ and $\lambda := \langle \Lambda \rangle$ yields
\begin{equation}
\label{eq::eom_aux_eff_nh_coupling}
	\begin{aligned}
		\dot{\alpha}_i(t) &= -\mathrm{i} (\omega_i - \mathrm{i}\gamma_i)\alpha_i(t) -\mathrm{i} g_i \lambda(t) \\
		\dot{\lambda}(t) &= -\mathrm{i} (\Omega -\mathrm{i} \Gamma) \lambda(t) - \mathrm{i} (g_1^* \alpha_1(t) + g_2^* \alpha_2(t)) \quad .
	\end{aligned}
\end{equation}
Formally integrating the second equation then leads to
\begin{equation}
  \lambda(t) = e^{-\mathrm{i}(\Omega -\mathrm{i} \Gamma) t} \lambda(0) - \mathrm{i} \int_0^t dt^\prime (g_1^* \alpha_1(t^\prime) + g_2^* \alpha_2(t^\prime) ) e^{-\mathrm{i}(\Omega - \mathrm{i} \Gamma)(t-t^\prime)}.
\end{equation}
We now focus on the case $\Gamma \gg \Omega, \omega_i, \gamma_i, g_i$ and try to approximate the integral in the equation above.
Therefore, terms like $\int_0^t dt^\prime g_i^* \alpha_i(t^\prime) e^{-\mathrm{i}(\Omega - \mathrm{i} \Gamma)(t-t^\prime)}$ are integrated by parts and terms where fractions of $\omega_i$, $\gamma_i$ or $g_i$ over $\Gamma$ occur are neglected.
\begin{figure}[t]
\centering
  \includegraphics[width=0.75\textwidth]{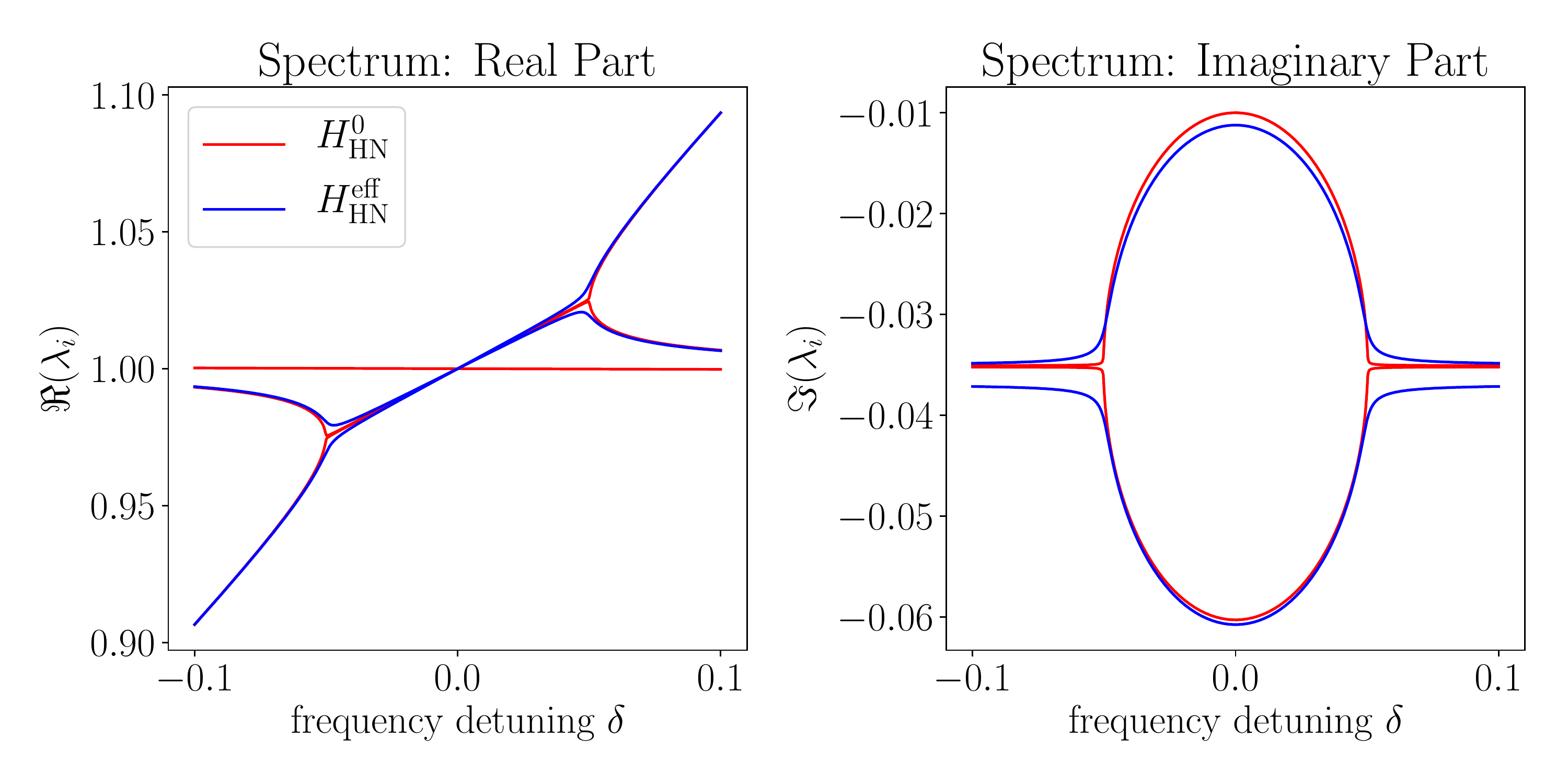}
  \caption{Real part and imaginary part of the eigenvalues $\lambda_i$ of $H_\mathrm{NH}^0$ and $H_\mathrm{NH}^\mathrm{eff}$ as a function of the frequency detuning $\delta = \omega_1 - \omega_2$ for the sample values $\omega_1 = 1.0 + \delta$, $\omega_2 = \Omega = 1.0$, $\gamma_1 = 0.01$, $\gamma_2 = 0.012$, $\Gamma = 10.0$ and $g_1 = g_2 = 0.5$.}
  \label{fig::eff_ham_spec_comp}
\end{figure}
The simplified expression for $\lambda(t)$ can then be inserted into the equations for $\dot{\alpha}_i$ in Eq.~(\ref{eq::eom_aux_eff_nh_coupling}) resulting in two equations which are completely decoupled from the auxilary mode
\begin{equation}
\begin{aligned}
	\dot{\alpha}_i &= -\mathrm{i} (\omega_i - \mathrm{i} \gamma_i) \alpha_i - \mathrm{i} g_i \left( e^{-\mathrm{i}(\Omega - \mathrm{i} \Gamma )t} \lambda(0) - \frac{1}{\Omega - \mathrm{i}\Gamma} (g_1^* \alpha_1(t) - \alpha_1(0) e^{-\mathrm{i}(\Omega - \mathrm{i}\Gamma) t} )  + g_2^* (\alpha_2(t) -\alpha_2(0) e^{-\mathrm{i} (\Omega - \mathrm{i}\Gamma) t} \right) \approx \\
	& \approx -\mathrm{i}\left( (\omega_i - \frac{|g_i|^2 \Omega}{\Omega^2 + \Gamma^2}) - \mathrm{i} (\gamma_i + \frac{|g_i|^2 \Gamma}{\Omega^2 + \Gamma^2}) \right) \alpha_i + \mathrm{i} g_i \frac{\Omega + \mathrm{i} \Gamma}{\Omega^2 + \Gamma^2} g_{i-1}^* \alpha_{i-1} \quad .
\end{aligned}
\end{equation}
Thus, the same dynamics of the complex amplitudes $\alpha_i$ is obtained by considering the effective Hamiltonian
\begin{equation}
  H_\mathrm{NH}^\mathrm{eff} = (\omega_1^\mathrm{eff} - \mathrm{i}\gamma_1^\mathrm{eff}) a_1^\dagger a_1 + (\omega_2^\mathrm{eff} - \mathrm{i}\gamma_2^\mathrm{eff}) a_2^\dagger a_2 + g_{1,2}^\mathrm{eff} a_1^\dagger a_2 + g_{2, 1}^\mathrm{eff} a_2^\dagger a_1
\end{equation}
with non Hermitian coupling strengths
\begin{equation}
  	g_{i, i+1}^\mathrm{eff} = - g_i \frac{\Omega + \mathrm{i} \Gamma}{\Omega^2 + \Gamma^2} g_{i+1}^* \approx - \mathrm{i}   \frac{g_i g_{i+1}^*
}{\Gamma}
\end{equation}
and renormalized frequencies and coupling strengths
\begin{equation}
  \begin{aligned}
  	\omega_i^\mathrm{eff} &= \omega_i - \frac{|g_i|^2 \Omega}{\Omega^2 + \Gamma^2} \approx \omega_i \\
  	\gamma_i^\mathrm{eff} &= \gamma_i + \frac{|g_i|^2 \Gamma}{\Omega^2 + \Gamma^2} \approx \gamma_i + \frac{|g_i|^2}{\Gamma} \quad.
  \end{aligned}
\end{equation}
We can check that by comparing the spectra of $H_\mathrm{NH}^0$ and $H_\mathrm{NH}^\mathrm{eff}$ as a function of the frequency detuning $\delta = \omega_1 - \omega_2$ for the sample values $\omega_1 = 1.0 + \delta$, $\omega_2 = \Omega = 1.0$, $\gamma_1 = 0.01$, $\gamma_2 = 0.012$, $\Gamma = 10.0$ and $g_1 = g_2 = 0.5$.
As shown in Fig.~\ref{fig::eff_ham_spec_comp}, the approximation works quite well independently of $\delta$.
Therefore, non-Hermitian interactions can be used to effectively describe two modes which are indirectly coupled via a lossy auxiliary mode.

\subsection{Derivation of equations of motion for non-Hermitian coupled bosonic modes}

In the case of our system where a central, driven, auxiliary mode is coupled non-Hermitianly to $\mathcal{N}$ main modes, the commutator in Eq.~(\ref{eq:nh_langevin}) reads
\begin{equation}
  [H_\mathrm{NH}, a_i] = \left\{\begin{array}{cc}
 	-(\omega_0 - \mathrm{i} \gamma_0) a_0 - \frac{g}{\sqrt{\mathcal{N}}} e^{\mathrm{i}\Theta} \sum_{j=1}^\mathcal{N} a_j - \mathrm{i}\eta e^{-\mathrm{i}\Omega t} & i = 0 \\
 	-(\omega - \mathrm{i} \gamma) a_i - g e^{\mathrm{i}\Theta} a_0 & i = 1, \ldots, \mathcal{N}
 \end{array}\right.
 \end{equation}
Assuming a bath at zero temperature, the dynamics of the expectation values $\alpha_i = \langle a_i \rangle$ in the co-rotating frame is
\begin{equation}
  \begin{aligned}
  	\dot{\alpha}_0 &= -\mathrm{i} (\omega_0 - \Omega -\mathrm{i} \gamma_0) \alpha_0 - \mathrm{i} \frac{g}{\sqrt{\mathcal{N}}} e^{\mathrm{i}\Theta} \sum_{j = 1}^\mathcal{N} \alpha_j +\eta \\
  	\dot{\alpha}_i &= -\mathrm{i} (\omega - \Omega -\mathrm{i} \gamma) \alpha_i - \mathrm{i} \frac{g}{\sqrt{\mathcal{N}}} e^{\mathrm{i}\Theta} \alpha_0
  \end{aligned}
\end{equation}
thus, leading to the matrix-vector form
\begin{equation}
\label{eq::eom_ampl}
  \dot{\mathbf{A}} = -\mathrm{i} \mathcal{H} \mathbf{A} +\eta (1, 0, \ldots)^T
\end{equation}
with $\mathbf{A} = (\alpha_0, \alpha_1, \ldots, \alpha_\mathcal{N})^T$ and
\begin{equation}
  \mathcal{H} =  \left(
  \begin{array}{cccc}
  \delta_0 - \mathrm{i} \gamma_0 & \frac{g}{\sqrt{\mathcal{N}}} e^{\mathrm{i}\Theta} & \ldots & \frac{g}{\sqrt{\mathcal{N}}}e^{\mathrm{i}\Theta}\\
  \frac{g}{\sqrt{\mathcal{N}}} e^{\mathrm{i}\Theta} & \delta - \mathrm{i} \gamma &  &    \\
  \vdots &  & \ddots &  \\
  \frac{g}{\sqrt{\mathcal{N}}} e^{\mathrm{i}\Theta} &  &  & \delta - \mathrm{i} \gamma
  \end{array}\right) \quad,
\end{equation}
where $\delta_{(0)} = \omega_{(0)} - \Omega$ (for the case $\eta=0$, set $\Omega = 0$, so that $\delta = \omega$).
This set of $\mathcal{N}+1$ linear, complex-valued equations can be brought into a system of $2(\mathcal{N}+1)$, nonlinear equations by separating the complex amplitudes $\alpha_i$ into real-values amplitudes and phases, i. e. $\alpha_i = r_i e^{\mathrm{i}\phi_i}$ with $r_i \equiv |\alpha_i|$, $\phi_i \equiv \arg{\alpha_i}$.
The derivative on the left side of Eq.~(\ref{eq::eom_ampl}) reads $\dot{\alpha}_i = \dot{r}_i e^{\mathrm{i}\phi_i} + \mathrm{i} \dot{\phi}_i \, r_i e^{\mathrm{i}\phi_i}$ so that by replacing $\alpha_i$ by $r_i e^{\mathrm{i}\phi_i}$ both sides can be decomposed into real and imaginary parts.
This yields two sets of coupled equations, $\mathcal{N}+1$ equations determining the evolution of the real-valued amplitudes
\begin{equation}
  \begin{aligned}
  	\dot{r}_0 &= -\gamma_0 r_0 + \frac{g}{\sqrt{\mathcal{N}}} \sum_{j=1}^\mathcal{N} \sin{(\Theta + \phi_j - \phi_0)} r_j + \eta \cos{\phi_0} \\
  	\dot{r}_i &= -\gamma r_i + \frac{g}{\sqrt{\mathcal{N}}} \sin{(\Theta + \phi_0 - \phi_i)}r_0 + \eta \cos{\phi_i},\quad i=1, \ldots, \mathcal{N}
  \end{aligned}
\end{equation}
and $\mathcal{N}+1$ equations for the phases
\begin{equation}
  \begin{aligned}
  	\dot{\phi}_0 &= - \delta_0 - \frac{g}{\sqrt{\mathcal{N}}} \sum_{j=1}^\mathcal{N} \frac{r_j}{r_0} \, \cos{(\Theta + \phi_j - \phi_0)} - \eta \sin{\phi_0} \\
  	\dot{\phi}_i &= - \delta - \frac{g}{\sqrt{\mathcal{N}}}  \frac{r_0}{r_i} \, \cos{(\Theta + \phi_0 - \phi_i)} - \eta \sin{\phi_i},\quad i=1, \ldots, \mathcal{N}
  \end{aligned}
\end{equation}
Apparently, we obtain a Kuramoto-like interaction for $\Theta = \pi/2$.
If we assume that the ratio $r_i / r_j$ is nearly constant after a sufficiently long settling time (as apparent from numerical simulations), the equation for the phases decouples from the amplitude equations and a Kuramoto model with a $\mathcal{N}$-to-$1$ coupling is effectively obtained.
Since there is an analytical solution for the Kuramoto model with an all-to-all coupling in the limit $\mathcal{N} \rightarrow \infty$, we present next a basis transformation that introduces all-to-all coupled, collective modes thus leading to the original Kuramoto model.

\section{Phase synchronization}
\label{sec::APNDX2}
\subsection{Mapping onto the Kuramoto model}

In order to map the non-Hermitian synchronization model onto an analytically solvable all-to-all coupled Kuramoto model, we first diagonalize the evolution matrix $\mathcal{H}$ and then find a transformation which turns $\mathcal{H}$ into another matrix $\mathcal{M}$ which possesses entries coupling $\mathcal{N}$  of the total $\mathcal{N}+1$ modes to $\mathcal{N}$ modes with equal strength.
The remaining mode is an eigenmode which does not interact with any other mode.

Note that as $\mathcal{H}$ is non-Hermitian, there are sets of parameters $\delta_{(0)}, \gamma_{(0)}, g, \Theta$ which lead to a non-diagonalizable matrix.
Neverthess, one can show that in most cases we can simply assume that $\mathcal{H}$ is similar to a diagonal matrix $\mathcal{D}$.
The Fourier transformation $\mathcal{T}$ only acting on the $\mathcal{N} \times \mathcal{N}$ subspace of non-auxiliary modes
\begin{equation}
	\begin{aligned}
		\mathcal{T} &= \left(
  		\begin{array}{cc}
  			1 & 0 \\
  			0 & \mathcal{F}
  		\end{array}
  		\right) \\
  		\mathcal{F}_{i, j} &= \frac{1}{\sqrt{\mathcal{N}}} e^{\mathrm{i} \frac{2\pi i j}{\mathcal{N}}}, \quad i,j = 1, \ldots, \mathcal{N}
	\end{aligned}
\end{equation}
introduces from a physical point of view a so-called bright mode interacting with the auxiliary mode while $\mathcal{N}-1$ dark modes do not participate in system and remain isolated.
More precisely, the matrix $\mathcal{T} \mathcal{H} \mathcal{T}^{-1}$ reduces the task to an effective $2 \times 2$ diagonalization problem.
One can show that by the transformation $\mathcal{S}_\pm$,
\begin{equation}
  \mathcal{S}_\pm = \left( \begin{array}{ccc}
 	1 & \begin{array}{ccc} 0 & \ldots & 0 \end{array} & s_\pm \\
 		\begin{array}{c} 0 \\ \vdots \\ 0 \end{array} & \mathds{1}_{\mathcal{N}-1} & \begin{array}{c} 0 \\ \vdots \\ 0 \end{array} \\
 		-s_\pm & \begin{array}{ccc} 0 & \ldots & 0 \end{array} & 1
 \end{array}\right), \quad s_\pm = \frac{g e^{\mathrm{i}\Theta}}{\frac{\Delta \omega - \mathrm{i}\Delta \gamma}{2} \pm \mu} = - \frac{\frac{\Delta \omega - \mathrm{i} \Delta \gamma}{2} \mp \mu}{g e^{\mathrm{i}\Theta}}
\end{equation}
with $\Delta \omega = \omega_0 - \omega$, $\Delta \gamma = \gamma_0 - \gamma$ and $\mu = \sqrt{\left(\frac{\Delta \omega - \mathrm{i}\Delta \gamma}{2}\right)^2 + g^2 e^{2\mathrm{i}\Theta}}$ the final step towards the diagonal matrix $\mathcal{D}_\pm$ is accomplished.
The sign $\pm$ only denotes the order of eigenvalues on the diagonal.
In fact, one finds three different eigenvalues, thus obtaining
\begin{equation}
\label{eq::diagonal_matrix}
  \mathcal{D}_\pm = (\mathcal{S_\pm T}) \mathcal{H} (\mathcal{S_\pm T})^{-1} = \left( \begin{array}{ccccc}
 	\frac{\Delta \omega - \mathrm{i} \Delta \gamma}{2} \pm \mu & & \\
 	 &0&&& \\
 	 &&\ddots && \\
 	 &&& 0 & \\
 	 &&&& \frac{\Delta \omega - \mathrm{i} \Delta\gamma}{2} \mp \mu
 \end{array}\right) + (\delta - \mathrm{i}\gamma) \mathds{1}_{\mathcal{N}+1} \quad .
\end{equation}
It is now straightforward to see that
\begin{equation}
  \mathcal{V} = \left(
  \begin{array}{ccc}
  	1 & \begin{array}{ccc} 0 & \ldots & 0 \end{array} & 0 \\
  	\begin{array}{c} 0 \\ \vdots \\ 0 \end{array} & -\mathds{1}_{\mathcal{N}-1} & \begin{array}{c} 1 \\ \vdots \\ 1 \end{array} \\
  	0 & \begin{array}{ccc} 1 & \ldots & 1 \end{array} & 1
  \end{array}\right)
\end{equation}
yields the basis transformation we searched for, since
\begin{equation}
  \mathcal{M}_\pm = (\mathcal{VS_\pm T}) \mathcal{H} (\mathcal{VS_\pm T})^{-1} = \left( \begin{array}{cccc} \frac{\Delta \omega - \mathrm{i} \Delta \gamma}{2} \pm \mu & 0 & \ldots & 0 \\
  		0  & \frac{\Delta \omega - \mathrm{i} \Delta \gamma  \mp 2\mu}{2\mathcal{N}} & \ldots & \frac{\Delta \omega - \mathrm{i} \Delta \gamma \mp 2\mu}{2\mathcal{N}} \\
  		\vdots & \vdots & & \vdots \\
  		0 & \frac{\Delta \omega - \mathrm{i} \Delta \gamma  \mp 2\mu}{2\mathcal{N}} & \ldots & \frac{\Delta \omega - \mathrm{i} \Delta \gamma  \mp 2\mu}{2\mathcal{N}}
  		\end{array} \right) + (\delta -\mathrm{i} \gamma) \mathds{1}_{\mathcal{N}+1} \quad.
\end{equation}
The choice between the $M_+$ or $M_-$ version enables us to select one of the eigenvalues to be isolated whereas the remaining ones are mixed in the lower left block thereby introducing all-to-all couplings of equal strength scaling with $\mathcal{N}^{-1}$.

We can now transform Eq.~(\ref{eq::eom_ampl}) into the collective basis which we will also refer to as the Kuramoto basis.
With $\mathbf{P} = \mathcal{VST} \mathbf{A} = (\pi_0, \pi_1, \ldots, \pi_\mathcal{N})^T$, the equation of motion in the new basis is given by $\dot{\mathbf{P}} = -\mathrm{i}\mathcal{M}\mathbf{P} + \eta \mathcal{VST}(1, 0, \ldots)^T$.
In a component-wise notation, the equations read
\begin{equation}
  \begin{aligned}
  	\dot{\pi}_0 &= -\mathrm{i} (\delta_0^\mathrm{eff} - \mathrm{i} \gamma_0^\mathrm{eff}) \pi_0  + \eta\\
  	\dot{\pi}_i &= -\mathrm{i} (\delta^\mathrm{eff} - \mathrm{i} \gamma^\mathrm{eff})\pi_i - \mathrm{i} \frac{K_\mathrm{eff} e^{\mathrm{i}\Theta_\mathrm{eff}}}{\mathcal{N}}\sum_{j\neq i}^\mathcal{N} \pi_j -  \eta_\mathrm{eff} e^{\mathrm{i}\Theta_\mathrm{drive}}
  \end{aligned}
\end{equation}
with effective frequencies
\begin{equation}
\begin{aligned}
	\delta^\mathrm{eff}_0 - \mathrm{i} \gamma^\mathrm{eff}_0 &= \delta - \mathrm{i} \gamma + \frac{\Delta \omega - \mathrm{i} \Delta \gamma  \pm 2\mu}{2} \\
	\delta^\mathrm{eff} - \mathrm{i} \gamma^\mathrm{eff} &= \delta - \mathrm{i} \gamma + \frac{\Delta \omega - \mathrm{i} \Delta \gamma  \mp 2\mu}{2\mathcal{N}}
\end{aligned}
\end{equation}
 and coupling parameters
 \begin{equation}
  K_\mathrm{eff} e^{\mathrm{i} \Theta_\mathrm{eff}} = \frac{1}{2} \left( \Delta \omega -\mathrm{i} \Delta \gamma \mp 2\mu\right) \quad.
\end{equation}
Due to a complex prefactor in the transformation, the modes with $i=1, \ldots, \mathcal{N}$ are driven with strength $\eta_\mathrm{eff}$ and a phase shift $\Theta_\mathrm{drive}$ whereby
\begin{equation}
	\begin{aligned}
		\eta_\mathrm{eff} &= |s_\pm| \eta \\
		\Theta_\mathrm{drive} &= \arg{s_\pm} .
	\end{aligned}
\end{equation}
Finally, the separation into equations for phases $\psi_i := \arg{\pi_i} - \Theta_\mathrm{drive}$ and real amplitudes $\rho_i := |\pi_i|$ for the coupled modes leads to
\begin{equation}
\label{eq::phase_ampl_Kuramoto_picture}
  \begin{aligned}
  	\dot{\rho}_i &= -\gamma_\mathrm{eff} \rho_i + \frac{K_\mathrm{eff}}{\mathcal{N}} \sum_{j\neq i}^\mathcal{N} \rho_j \sin{(\Theta_\mathrm{eff} + \psi_j - \psi_i)} + \eta_\mathrm{eff} \cos{\psi_i} \\
  	\dot{\psi}_i &= -\delta_\mathrm{eff} - \frac{K_\mathrm{eff}}{\mathcal{N}} \sum_{j\neq i}^\mathcal{N} \frac{\rho_j}{\rho_i} \cos{(\Theta_\mathrm{eff} + \psi_j - \psi_i)} - \frac{\eta_\mathrm{eff}}{\rho_i} \sin{\psi_i}\quad, i=1, \ldots, \mathcal{N}
  \end{aligned}
\end{equation}
The phase equations decouple from the amplitude equations after a sufficiently long-time, since firstly the long-term behaviour is dominated by the driving, which pumps each collective mode equally (except for the eigenmode of course).
Secondly, a steady state is reached, where $\rho_i = \rho = \text{const.}$ thus ensuring that the two equations decouple.
Also in the undriven case $\eta = 0$, the ratios $\rho_i / \rho_j \rightarrow 1$, since all collective modes are linear combinations of the eigenmodes given by $\mathcal{V}$.
A look at the diagonal matrix in Eq.~(\ref{eq::diagonal_matrix}) only one eigenmode survives in the long-time limit, corresponding to either the eigenvalue $(\Delta \omega -\mathrm{i}\Delta \gamma)/2 + \mu$ or $(\Delta \omega -\mathrm{i}\Delta \gamma)/2 - \mu$.
By choosing one of the transformations $\mathcal{S}_\pm$, we always can choose the fast decaying eigenmode to be the isolated mode in the new Kuramoto basis.
Then, for initial complex-amplitudes $\mathbf{A} = (\alpha_0(0), \alpha_1(0), \ldots)^T$ and a short notation for the eigenvalues $\lambda_0 = (\Delta \omega -\mathrm{i}\Delta \gamma)/2\pm \mu$, $\lambda_{0<j<\mathcal{N}}=\omega-\mathrm{i}\gamma$, $\lambda_\mathcal{N} = (\Delta \omega -\mathrm{i}\Delta \gamma)/2 \mp \mu$ one obtains
\begin{equation}
  \frac{\pi_i}{\pi_j} = \frac{\sum_k \mathcal{V}_{i, k} e^{-\mathrm{i}\lambda_k t}(\mathcal{S_\pm T}\mathbf{A}(0))_k}{\sum_k \mathcal{V}_{j, k} e^{-\mathrm{i}\lambda_k t} (\mathcal{S_\pm T}\mathbf{A}(0))_k} \rightarrow \frac{ \mathcal{V}_{i, \mathcal{N}} e^{-\mathrm{i}\lambda_\mathcal{N} t}(\mathcal{S_\pm T}\mathbf{A}(0))_\mathcal{N}}{\mathcal{V}_{j, \mathcal{N}} e^{-\mathrm{i}\lambda_\mathcal{N} t} (\mathcal{S_\pm T}\mathbf{A}(0))_\mathcal{N}} = 1, \quad i, j = 1, \ldots, \mathcal{N}
\end{equation}

We can therefore set $\rho_i / \rho_j \approx 1$ and concentrate on the phase equations which now clearly represent variations of the Kuramoto model with all-to-all couplings.
In particular, for $\Theta_\mathrm{eff} = \pi / 2$, the phases $\psi_i$ evolve according to
\begin{equation}
  \dot{\psi}_i = -\delta_\mathrm{eff} + \frac{K_\mathrm{eff}}{\mathcal{N}} \sum_{j\neq i}^\mathcal{N} \sin{(\psi_j - \psi_i)} - \frac{\eta_\mathrm{eff}}{\rho} \sin{\psi_i}\quad,
\end{equation}
which is the \emph{Kuramoto model} \cite{acebron2005thekuramoto} with driving.
In conclusion, the non-Hermitian dynamics can be mapped onto variations of the  Kuramoto model with all-to-all coupling.

\subsection{Synchronization in the Kuramoto-basis}

We showed, that in the long-time limit the phase equations for the Kuramoto modes are given by
\begin{equation}
\label{eq::sakaguchi_kuramot_driven}
  	\dot{\psi}_i = -\delta_\mathrm{eff} - \frac{K_\mathrm{eff}}{\mathcal{N}} \sum_{j\neq i}^\mathcal{N} \cos{(\Theta_\mathrm{eff} + \psi_j - \psi_i)} - \frac{\eta_\mathrm{eff}}{\rho} \sin{\psi_i}\quad.
\end{equation}
We can now investigate synchronization behaviour of this reduced model by taking advantage of existing solutions.
We first start with the undriven case $\eta = 0$ and assume large $\mathcal{N}$ since many exact results we cite rely on this requirement.
However, the key message also holds for small $\mathcal{N}$.

In case of $\eta = 0$, Eq.~(\ref{eq::sakaguchi_kuramot_driven}) is known as the \emph{Sakaguchi-Kuramoto model} for which an analytical solution is provided.
In the case of a frequency distribution $g(\omega) = \delta(\omega - \omega_\mathrm{eff})$, synchronization is expected to occur for any $K_\mathrm{eff} > 0$.
More importantly, in the final, stationary state all phases $\psi_i$ are equal, i. e. the synchronization order parameter $z$ defined by $ze^{\mathrm{i}\Psi} := \sum_i e^{\mathrm{i}\psi_i}$ is $z = 1$.
Consequently, the mean frequency is $\bar{\omega}= - \omega_\mathrm{eff} - K_\mathrm{eff} \cos{\Theta_\mathrm{eff}}$.

Apparently, in the Kuramoto basis all coupled modes synchronize while the remaining eigenmode decays significantly faster, and therefore the amplitude vector is approximately
\begin{equation}
  \mathbf{P}(t) \rightarrow \left(\begin{array}{c}
  	0 \\
  	\rho e^{\mathrm{i}\Psi(t)} \\
  	\vdots \\
  	\rho e^{\mathrm{i}\Psi(t)}
  \end{array}\right) \quad.
\end{equation}
If the back-transformation $\mathbf{A} = (\mathcal{VS_\pm T})^{-1} \mathbf{P}(t)$ is supposed to protect the synchronization behaviour, the phase of all $\alpha_i$ must be the same.
The special property of $(\mathcal{VS_\pm T})^{-1}$,
\begin{equation}
  (\mathcal{VS_\pm T})^{-1} \left(\begin{array}{c}
  	0\\1\\ \vdots \\ 1
  \end{array}\right) \propto \left(\begin{array}{c}
  	-\sqrt{\mathcal{N}}  s_\pm\\1\\ \vdots \\ 1
  \end{array}\right)
\end{equation}
then leads to synchronization conditions, since $s_\pm$ must be real.
Otherwise the phase of the auxiliary mode $\alpha_0$ would be different from the other phases.
One can show that $s_\pm \in \mathbb{R}$ is satisfied if
\begin{equation}
  \tan{\Theta} = -\frac{\Delta \gamma}{\Delta \omega} \quad ,
\end{equation}
which is differently derived in the subsequent section.
To further rule out the case of a phase shift of $\pi$ between the auxiliary mode and the main modes, $s_\pm < 0$ is required, which leads to the additional condition $\Delta \gamma \sin{\Theta} < \Delta \omega \cos{\Theta}$.

The synchronization conditions can be derived in a similar way for the driven case $\eta \ne 0$.
Eq. (\ref{eq::sakaguchi_kuramot_driven}) is solved for $\psi_i = \psi_j \equiv \psi$ and a steady state amplitude $\rho_i = \rho$ given by
\begin{equation}
  	\rho e^{\mathrm{i}\psi} = \frac{\mathrm{i}s_+ \eta}{\delta - \mathrm{i}\gamma + \frac{\Delta \omega -\mathrm{i}\Delta \gamma - 2\mu}{2}}
\end{equation}
where the $+$-version of the transformation $(\mathcal{VS_\pm T})$ was chosen.
The steady state of the isolated mode is
\begin{equation}
  \rho_0 e^{\mathrm{i}\psi_0} = -\frac{\mathrm{i}\eta}{\delta -\mathrm{i} \gamma + \frac{\Delta \omega -\mathrm{i}\Delta\gamma + 2\mu}{2}}
\end{equation}
and by performing the back-transformation $\mathbf{A} = (\mathcal{VS_+ T})^{-1} \mathbf{P}$ again,
one finds that the phases remain synchronized in the bare basis only if the synchronization conditions
\begin{equation}
  \begin{aligned}
  	\tan{\Theta} &= -\frac{\gamma}{\omega - \Omega} \\
  	(\omega - \Omega) \cos{\Theta} &< \gamma \sin{\Theta}
  \end{aligned}
\end{equation} are satisfied.
In fact, these conditions ensure that the relative phases of the modes are unaffected by the transformation $\mathcal{VS_+ T}$.

\subsection{Derivation of synchronization conditions for non-Hermitianly coupled modes}

The system is synchronized if $\left.\Im{\left(\alpha_j / \alpha_0\right)}\right|_{t \rightarrow \infty} = 0$ and $\left.\Re{\left(\alpha_j / \alpha_0\right)}\right|_{t \rightarrow \infty} > 0$ since then $\phi_j = \phi_0$ for $t \rightarrow \infty$.
To calculate those ratios of complex amplitudes, Eq.~(\ref{eq::eom_ampl}) is solved by diagonalizing $\mathcal{H} = (\mathcal{S_\pm T})^{-1}\mathcal{D_\pm}(\mathcal{S_\pm T})$,
\begin{equation}
\label{eq::compl_eom_solution}
  \mathbf{A}(t) = (\mathcal{S_\pm T})^{-1} e^{-\mathrm{i}\mathcal{D_\pm}(t - t_0)}(\mathcal{S_\pm T}) \mathbf{A}(t_0) + \eta \int_{t_0}^t dt^\prime (\mathcal{S_\pm T})^{-1} e^{-\mathrm{i}\mathcal{D_\pm}(t - t^\prime)}(\mathcal{S_\pm T}) \left(\begin{array}{c}
  		1 \\ 0 \\ \vdots \\ 0
  	\end{array}\right) \quad ,
\end{equation}
with $t_0=0$.

We start with the case $\eta = 0$.
In order to find $\alpha_i$, $\alpha_0$ the matrices-vector products in Eq.~(\ref{eq::compl_eom_solution}) are carried out leading to
\begin{equation}
\begin{aligned}
	\alpha_j(t) &= e^{-\mathrm{i} (\omega-\mathrm{i}\gamma) t}  \sum_{l=0}^\mathcal{N} \alpha_l(0) \Biggl[ (\mathcal{S_\pm T})^{-1}_{j, 0} (\mathcal{S_\pm T})_{0, l} e^{-\mathrm{i}\left(\frac{\Delta \omega -\mathrm{i}\Delta \gamma}{2}\pm\mu\right) t } + \\
	&+ \sum_{k=0}^{\mathcal{N}-1} (\mathcal{S_\pm T})^{-1}_{j, k} (\mathcal{S_\pm T})_{k, l} + \\
	& + (\mathcal{S_\pm T})^{-1}_{j, \mathcal{N}} (\mathcal{S_\pm T})_{\mathcal{N}, l} e^{-\mathrm{i}\left(\frac{\Delta \omega -\mathrm{i}\Delta \gamma}{2}\mp\mu\right) t }  \Biggr] \quad.
\end{aligned}
\end{equation}
Considering the infinite-time limit $t \rightarrow \infty$, the first two lines are neglected under the assumption that w. l. o. g. $(\mathcal{D}_\pm)_{\mathcal{N}\mathcal{N}} = \frac{\Delta \omega -\mathrm{i}\Delta \gamma}{2}\mp\mu$ is the long-living eigenvalue with the smallest imaginary part.
Therefore, for $t \rightarrow \infty$ the last term dominates the other ones and we can approximately write
\begin{equation}
  \frac{\alpha_j}{\alpha_0} \approx \frac{e^{-\mathrm{i} (\omega-\mathrm{i}\gamma) t}  \sum_{l=0}^\mathcal{N} \alpha_l(0) (\mathcal{S_\pm T})^{-1}_{j, \mathcal{N}} (\mathcal{S_\pm T})_{\mathcal{N}, l} e^{-\mathrm{i}\left(\frac{\Delta \omega -\mathrm{i}\Delta \gamma}{2}\mp\mu\right) t }}{e^{-\mathrm{i} (\omega-\mathrm{i}\gamma) t}  \sum_{l=0}^\mathcal{N} \alpha_l(0) (\mathcal{S_\pm T})^{-1}_{0, \mathcal{N}} (\mathcal{S_\pm T})_{\mathcal{N}, l} e^{-\mathrm{i}\left(\frac{\Delta \omega -\mathrm{i}\Delta \gamma}{2}\mp\mu\right) t }} = \frac{(\mathcal{S_\pm T})^{-1}_{j, \mathcal{N}}}{(\mathcal{S_\pm T})^{-1}_{0, \mathcal{N}}} = -\frac{1}{\sqrt{\mathcal{N}} s_\pm} \quad .
\end{equation}
Taking the imaginary and real part as well as inserting the definition of $s_\pm$ then finally leads to the synchronization conditions for the undriven case
\begin{equation}
  \tan{\Theta} = -\frac{\Delta \gamma}{\Delta \omega}
\end{equation}
and
\begin{equation}
  \Delta \gamma \sin{\Theta} < \Delta \omega \cos{\Theta}
\end{equation}

The derivation for finite $\eta \ne 0$ is in principle similar.
In contrast to the previous case, the term surviving for $t\rightarrow \infty$ is the integral term in Eq.~ (\ref{eq::compl_eom_solution}) and one can write
\begin{equation}
  \alpha_k(t\rightarrow \infty) = \int_0^t dt^\prime \sum_{l=0}^\mathcal{N} (\mathcal{S_\pm T})^{-1}_{k, l} e^{-\mathrm{i} (\mathcal{D}_\pm)_{l, l} (t - t^\prime)} (\mathcal{S_\pm T})_{l, 0} \rightarrow \mathrm{i} \sum_{l=0}^{\mathcal{N}} \frac{(\mathcal{S_\pm T})^{-1}_{k, l} (\mathcal{S_\pm T})_{l, 0}}{(\mathcal{D}_\pm)_{l, l}} \quad.
\end{equation}
By inserting the corresponding expressions for the matrix elements and some mathematical reordering of terms one eventually finds that the ratio of complex amplitudes approach a constant value
\begin{equation}
  \frac{\alpha_k}{\alpha_0} \rightarrow -\frac{1}{\sqrt{\mathcal{N}}} \frac{g e^{\mathrm{i} \Theta}}{\delta - \mathrm{i} \gamma}, \quad t \rightarrow \infty \quad.
\end{equation}
By requiring the imaginary part of the expression above to be zero the synchronization conditions
\begin{equation}
  \tan{\Theta} = -\frac{\gamma}{\omega - \Omega}
\end{equation}
and
\begin{equation}
  \cos{\Theta} (\omega - \Omega) < \sin{\Theta} \gamma
\end{equation}
are obtained.

\section{Dynamics in the presence of external noise}
\label{sec::APNDX3}

In order to test the resilience of synchronized systems against noise, we consider the system now to be exposed to classical, thermal noise.
To this end, the complex amplitudes $\alpha_i$ are separated into real parts $q_i, p_i$, $\alpha_i = q_i + \mathrm{i} p_i$ resembling position and momentum coordinates.
We can then write down two equations of motion for $q_i$ and $p_i$ and include thermal forces modelled by a Wiener process which is added to the equation for $p_i$
\begin{equation}
  \begin{aligned}
  	d{q}_0 &= \left[ \omega_0 p_0 - \gamma_0 p_0 + \frac{g}{\sqrt{\mathcal{N}}} \sum_{i = 1}^\mathcal{N} \left(\cos{\Theta} \, p_i + \sin{\Theta} \, q_i \right)\right] dt \\
  	d{q}_i &= \left[ \omega p_i - \gamma q_i + \frac{g}{\sqrt{\mathcal{N}}} \left(\cos{\Theta} \,  p_0 + \sin{\Theta} \, q_0 \right) \right] dt \\
  	d{p}_0 &= \left[ -\omega_0 q_0 - \gamma_0 p_0 + \frac{g}{\sqrt{\mathcal{N}}} \sum_{i = 1}^\mathcal{N} \left( \sin{\Theta} \, p_i - \cos{\Theta} \, q_i \right) \right] dt + dW_0(t) \\
  	d{p}_i &= \left[ -\omega q_i - \gamma p_i + \frac{g}{\sqrt{\mathcal{N}}} \left( \sin{\Theta} \, p_0 - \cos{\Theta} \, q_0 \right) \right] dt + dW_i(t) \quad ,
  \end{aligned}
\end{equation}
where $dW_i$ are mutually independent Wiener increments.
These obey
\begin{equation}
  \begin{aligned}
  	\langle dW_i(t) \rangle &= 0 \\
  	\langle dW_i(t) dW_j(t) \rangle &= \delta_{i, j}\, \xi_i \xi_j \, dt
  \end{aligned}
\end{equation}
with $\xi_i = \sqrt{2 \gamma_i n_i(T)}$ and the occupation $n_i(T) = \frac{k_B T}{\hbar \omega_i}$ ($\gamma_{i>0}=\gamma$, $\omega_{i>0}=\omega$).
To obtain the (stochastic) equation of motion for the complex amplitudes, the equations for $q_i$ and $p_i$ are combined again, yielding
\begin{equation}
\label{eq::compl_ampl_sde}
  d\mathbf{A} = \left(-\mathrm{i} \mathcal{H} \mathbf{A} + \eta \mathbf{u} \right) dt + \mathrm{i} d\mathbf{W} \quad.
\end{equation}
Here, the vector matrix notation was extended by the Wiener-noise vector $d\mathbf{W} = (dW_0, dW_1, \ldots, dW_\mathcal{N})^T$.

For numerical simulations, Eq.~(\ref{eq::compl_ampl_sde}) was solved numerically, by integration for randomly drawn, finite Wiener increments.
More precisely, the formal solution of Eq.~(\ref{eq::compl_ampl_sde})
\begin{equation}
  \mathbf{A}(t) = \exp{\left(-\mathrm{i} \mathcal{H} t\right)} \mathbf{A}(0) + \eta \mathbf{u} \int_0^t dt^\prime \exp{\left(-\mathrm{i} \mathcal{H} (t - t^\prime)\right)} + \mathrm{i} \int_0^t  \exp{\left(-\mathrm{i}\mathcal{H}(t - t^\prime)\right)} d\mathbf{W}(t^\prime) \quad ,
\end{equation}
is used to approximate the last integral as a sum of finite Wiener increments $\Delta \mathbf{W}$ drawn from a normal distribution with a width proportional to the square root of the step size $\Delta t$ of the time discretization.
By collecting the multiple random, sample paths, averages and variances are easily calculated for several quantities such as real-valued amplitudes and phase differences.

\end{document}